# Model-based Joint Analysis of Safety and Security: Survey and Identification of Gaps*

Stefano M. Nicoletti [a], Marijn Peppelman [a], Christina Kolb [b], Mariëlle Stoelinga [a,c]

[a] *University of Twente, Formal Methods and Tools, Enschede, The Netherlands*
[b] *University of Twente, Industrial Engineering & Business Information Systems, Enschede, The Netherlands*
[c] *Radboud University, Department of Software Science, Nijmegen, The Netherlands*

**Abstract**

We survey the state-of-the-art on model-based formalisms for safety and security joint analysis, where safety refers to the absence of unintended failures, and security to absence of malicious attacks. We conduct a thorough literature review and — as a result — we consider fourteen model-based formalisms and compare them with respect to several criteria: (1) *Modelling capabilities and Expressiveness:* which phenomena can be expressed in these formalisms? To which extent can they capture safety-security interactions? (2) *Analytical capabilities:* which analysis types are supported? (3) *Practical applicability:* to what extent have the formalisms been used to analyze small or larger case studies? Furthermore, (1) we present more precise definitions for safety-security dependencies in tree-like formalisms; (2) we showcase the potential of each formalism by modelling the same toy example from the literature and (3) we present our findings and reflect on possible ways to narrow highlighted gaps. In summary, our key findings are the following: (1) the majority of approaches combine tree-like formal models; (2) the exact nature of safety-security interaction is still ill-understood and (3) diverse formalisms can capture different interactions; (4) analyzed formalisms merge modelling constructs from existing safety- and security-specific formalisms, without introducing *ad hoc* constructs to model safety-security interactions, or (5) metrics to analyze trade offs. Moreover, (6) large case studies representing safety-security interactions are still missing.

---

*This work was partially funded by the NWO grant NWA.1160.18.238 (PrimaVera), and the European Union's Horizon 2020 research and innovation programme under the Marie Skłodowska-Curie grant agreement No 101008233, and the ERC Consolidator Grant 864075 (*CAESAR*).

*Email addresses:* `s.m.nicoletti@utwente.nl` (Stefano M. Nicoletti), `m.peppelman@utwente.nl` (Marijn Peppelman), `c.kolb@utwente.nl` (Christina Kolb), `m.i.a.stoelinga@utwente.nl` (Mariëlle Stoelinga)

## 1. Introduction

New technology comes with new risks: self-driving cars or train automation systems [72] may get hacked, people depend on the proper functioning of medical implants for their continued health [90]. Such risks concern both accidental failures (safety) and malicious attacks (security). Safety and security can be heavily intertwined. Measures that increase safety may decrease security and vice versa: the Internet-of-Things offers ample opportunities to monitor the safety of a power plant, but their many access points are notorious for enabling hackers to enter the system [49]. Passwords secure patients' medical data, but are a hindrance during emergencies. This is not an entirely new problem [32]: in fact, it has been widely acknowledged — including by international risk standards [40, 41] — that safety and security must be analyzed in combination [11, 64]. To cater for this need, various risk frameworks have been developed for the joint analysis of safety and security, i.e. the safety-security co-analysis. *Process-oriented* frameworks consider the steps needed for performing safety-security risk analysis [66]. Various formalisms specifically support the *risk assessment phase* within this process, i.e., the identification, analysis and evaluation of safety-security risks. Text-based methods include FMVEA [79, 80], CHASSIS [79, 70], HAZOP [55, 31] and SAHARA [58].

**Motivation.** Our focus, however, is on *model-based* risk assessment. Model-based formalisms provide a detailed insight in different failure and attack modes, mechanisms and their root causes. Understanding these is crucial for effective decision making. Furthermore, model-based formalisms allow one to compute various dependability metrics: such metrics quantify key performance indicators, such as the system availability and mean time to attack/failure. The goal of this paper is to survey the state-of-the-art on these model-based formalisms for safety-security co-analysis, as an overarching work unitedly addressing this literature space is still lacking. We compare these formalisms with respect to several criteria: (1) *Modelling capabilities and Expressiveness:* which phenomena can be expressed in these formalisms? To which extent can they capture safety-security interactions? (2) *Analytical capabilities:* which analysis types are supported? (3) *Practical applicability:* to what extend have the formalisms been used to analyze small or larger case studies? These questions summarize well known criteria used when selecting appropriate formal methods for a given task, and synthesise principles seen in corresponding literature [61].

**Methodology: an overview.** To address the proposed task, we conducted a systematic and standard literature review process, which yielded 14 formalisms for safety-security co-analysis (see Table 1). We grouped these into 4 categories: *Combinations of attack trees and fault trees*; *Formalisms extending attack trees and/or fault trees*; *Mathematical formalisms* and *Architectural formalisms* that extend architectural system models with additional means for safety-security aspects. For *expressiveness*, we compare to what extent these formalisms are able to capture the four safety-security interactions identified by Kriaa et al. [53]: *Conditional dependency*, where security requirements necessitate safety requirements, or vice-versa; *Mutual reinforcement*, where safety requirements or measures increase security, or vice-versa; *Antagonism*, where safety and security requirements or measures conflict with each other; and *Independence*, where no interaction takes place. To illustrate what the formalisms look like, we model in each formalism the Locked Door Example [85, 51]. This is a classical example of safety-security interaction, exemplifying an antagonistic dependency between safety and security: if locked, a door is unsafe in case of a fire. However, if unlocked, the door is insecure in case of a burglary. Our methodology is thoroughly presented in Sec. 2.

**Key findings.** Our survey revealed several noteworthy findings, some of which are summarized in Table 2 and Table 3. First, most formalisms are based on extensions of Attack Trees (ATs) and Fault Trees (FTs). Since FTs and ATs are widespread formalisms that model how systems can fail or be attacked, this is not surprising. Extensions of these formalisms are Attack-Fault Trees (AFTs) [9], Component Fault Trees (CFTs) [44, 83], Fault Tree/Attack Trees (FT/ATs) [35], Boolean driven Markov processes (BDMPs) [15], Attack Tree Bow Ties (ATBTs) [3, 2], Failure-Attack-CounTermeasure (FACT) Graphs [77], and State/Event Fault Treess (SEFTs) [73]. Secondly, the investigated formalisms can capture different safety-security interactions. These formalisms are mostly based on safety-specific or security-specific techniques, that are further adapted to account for the other. However, none of the formalisms introduce novel modeling constructs to capture safety-security interactions, and neither are novel metrics introduced to capture dependencies between safety and security. Moreover, some methods can only provide qualitative insight into the safety-security relationships, while others can also account for quantitative (failure/attack) probabilities. Furthermore, a large-sized industrial case study has not yet been conducted. Finally, some of the investigated modelling methods could be enhanced to increase their expressiveness. We detail these findings and reflect on highlighted gaps in Sec. 10.

| Formalism | Ref. | Year | #Citations |
|---|---|---|---|
| Fault Tree/Attack Tree Integration (FT/AT) | [35] | 2009 | 192 |
| Component Fault Trees (CFTs) | [83] | 2013 | 65 |
| Attack-Fault Trees (AFTs) | [54] | 2017 | 86 |
| State/Event Fault Trees (SEFTs) | [73] | 2013 | 32 |
| Failure-Attack-CounTermeasure (FACT) Graphs | [77] | 2015 | 85 |
| Boolean Driven Markov Processes (BDMPs) | [51] | 2014 | 45 |
| Attack Tree Bow-ties (ATBTs) | [3] | 2018 | 151 |
| Bayesian Networks (BNs) | [48] | 2015 | 56 |
| Threats-Hazards-Opportunities (THO) Framework | [10] | 2007 | 367 |
| STAMP | [38] | 2017 | 197 |
| SysML | [67] | 2011 | 92 |
| ALLOY | [42] | 2002 | 1678 |
| Event-B | [89] | 2017 | 16 |
| Architectural Analysis and Design Language (AADL) | [28] | 2020 | 2 |

Table 1: Overview of safety-security formalisms. Citations from Google Scholar, October 2022.

**Related work.** There are a few earlier surveys [59] on safety-security analysis methods. An important survey was published by Kriaa et al. [53] that introduces definitions for safety and security interactions. The authors compare non-model-based and model-based approaches, the latter being graphical or non-graphical. The survey [26] compares seven frameworks with respect to model creation, origin, stages of the risk assessment, and main use cases. The comparison conducted here is, however, carried out at a high level. The survey [65] summarizes current practices in safety and in security modeling. Then a new model is proposed, combining the goal structuring notation GSL with Attack-Defense Trees. In contrast to their work, we focus on model-based approaches and we present more detailed definitions for safety and security interactions. Furthermore, we employ a running example that we model in every analyzed formalism whenever possible. We then compare formalisms with respect to their ability to model safety-security interactions. Our work details how the formalisms work, what their technical capabilities are, and how safety-security interact. The survey [60] provides an overview about text-based models for engineering. It investigates research questions such as at which development stage the research was conducted, which methods and tools were employed during the research — e.g., FMVEA [79, 80], CHASSIS [79, 70], HAZOP [55, 31] and SAHARA [58] — what the classification of the contribution of the research is, in which research domains were the results evaluated, where was the research published, and what the research publication time-line and trend is. In contrast to their work, we focus our attention on model-based formalisms rather than on text-based ones, and we are interested in concrete dependencies between safety and security. Furthermore, we present examples for each formalism. Finally, the survey [71] provides an overview of different kinds of formalisms that consider safety or security, although separately. On the contrary, our focus is on formalisms that consider *both* safety and security.

| Formalism | Dependencies | | | | Expressiveness | | | Modelling | Application | Tool |
|---|---|---|---|---|---|---|---|---|---|---|
| | A | CD | MR | I | T | S | C | | | |
| **FT/AT** | $\overset{*}{\rightarrow}$ | $\rightarrow$ | | x | | | | ATs refine FT leaves | Chemical plant | |
| **CFTs** | $*$ | x | x | x | | | | Merge ATs + FTs | Cruise control | SafeTbox [88] |
| **AFTs** | x | x | x | x | x | | | Merge dynamic ATs + FTs | Pipeline, lock door | UPPAAL |
| **SEFTs** | x | $\rightarrow$ | x | x | | x | | FTs + Petri nets | Tyre pressure, lock door | ESSaRel |
| **FACT Graphs** | $\overset{*}{\rightarrow}$ | $\rightarrow$ | | x | x | | | FT/ATs + Triggers + Countermeasures | Overpressure vessel | |
| **BDMPs** | x | x | x | x | x | x | | Triggers, Petri nets | Pipeline, lock door | KB3, Figaro |
| **ATBTs** | $\overset{*}{\rightarrow}$ | $\rightarrow$ | | x | | | x | Bowties + FT/AT | Pipeline, Stuxnet | |
| **BNs** | x | x | x | x | $\circ$ | $\circ$ | x | Conditional prob. | Pipeline | MSBNx |
| **THO Framework** | $\overset{*}{\rightarrow}$ | $\rightarrow$ | | x | | | x | FTs + event trees | Electrical grid | |
| **STAMP** | | | | x | | | | Process controller | Synchronous-islanding | |
| **SysML** | | | | | | | | System components | Embedded systems | TTool |
| **ALLOY** | x | x | x | x | | x | | Prove system correctness | Fire detection system | ALLOY |
| **Event-B** | x | x | x | x | x | x | $\square$ | Prove system correctness | Charging system | RODIN |
| **AADL** | | | | x | | | | System components + ports | Lock door | Cheddar, Marzhin |

Table 2: Comparison of safety-security formalisms. A= Antagonism, CD= Conditional Dependency, MR=Mutual reinforcement, I=Independence. T = Time/order, S = States, C = Consequences. $*$ = capable when NOT-gate is supported. $\rightarrow$ = capable but only directional from security to safety. $\circ$ = only if BNs are dynamic. $\square$ = possible in principle, not showcased in literature.

**A summary of our contributions:**

**1)** We conduct a thorough literature review on model-based formalisms for joint analysis of safety and security.

**2)** We compare found formalisms by their modelling and analysis capabilities, their application domains in industry and expressiveness, i.e., their ability to capture safety-security dependencies.

**3)** We present more precise definitions for safety-security dependencies in tree-like formalisms (Sec. 6).

**4)** We showcase the potential of each formalism by modelling the same toy example from the literature [85, 51].

**5)** We present our findings and reflect on possible ways to narrow highlighted gaps (Sec. 10, Sec. 11).

| Formalism | Analysis | | Details |
|---|---|---|---|
| | QL | QT | |
| **FT/AT** | x | x | Same as base FTs (MCS, MPS, probabilities...). |
| **CFTs** | x | x | Same as base FTs (MCS, MPS, probabilities...) + extension on MCS analysis. |
| **AFTs** | x | x | Same as base FTs + time, cost, likelihood of an attack. Trade offs between attributes. |
| **SEFTs** | x | x | Same as FT/ATs plus details on the state of components via Petri nets. |
| **FACT Graphs** | x | x | Same as FT/ATs plus triggers and countermeasures |
| **BDMPs** | x | x | Mean Time to Success, probability of success, list of possible attack success sequences. |
| **ATBTs** | x | x | Risk level evaluation like regular BTs, trade offs analysis. |
| **BNs** | x | x | Calculate reliability metrics (mean time to failure) and conditional independence analysis. |
| **THO Framework** | x | x | Likelihood of an attack |
| **STAMP** | x | | Identify potential hazards and undesired behaviours. |
| **SysML** | x | | Checks for reachable states that violate safety properties/ security requirements. |
| **ALLOY** | x | | Prove properties of a system under given assumptions. |
| **Event-B** | x | | Prove properties of a system under given assumptions. |
| **AADL** | x | x | Calculate MTBFs of the system. Error model statements can generate FTs. |

Table 3: Overview of analysis capabilities of safety-security formalisms. QL=Qualitative Analysis, QT=Quantitative Analysis.

**Structure of the paper.** Sec. 2 details our methodology, Sec. 3 provides background for fault trees and attack trees., Sec. 4, 5, 7 and 8 compare the various formalisms. Sec. 6 presents our definitions for dependencies on tree-like formalisms, Sec. 9 presents a comparison of formalisms based on their expressiveness. Sec. 10 presents — and reflect on — our findings and Sec. 11 concludes the paper.

## 2. Methodology

**Research questions.** The goal of this survey is to understand the state-of-the-art on safety-security co-analysis and to identify future needs. In particular, when analyzing surveyed formalisms, we focus on the following questions: (**Q1**) How expressive are these formalisms, e.g., how many safety-security interaction can they capture? (**Q2**) Which modeling constructs exist to model the interactions between safety and security? (**Q3**) Which analyses do these formalisms enable? (**Q4**) How do these formalisms compare on industrial case studies and (**Q5**) What are the gaps and the findings? What would be desirable extensions? These questions summarize well known criteria employed to assess formal methods, and synthesise principles seen in corresponding literature [61].

**Search methodology.** We followed an established methodology for our literature search [18, 57]. Our aim was to find peer-reviewed publications from 1922 until 2020, focusing on model-based formalisms jointly analyzing safety and security. Following [18], we first queried the publication databases Scopus, Microsoft Academic and Google Scholar with relevant keywords. By using "safety" and "security" we collected 4997 results. We further refined them — 2714 entries — by focusing on "model-based" formalisms and subsequently by employing keywords such as "dependency" together with appropriate Boolean operators and wildcards. Finally, we refined our search by exploring the formalisms and tools we found (e.g., fault-attack trees, BDMP, AADL). We also explored earlier surveys on safety-security combinations [53, 65, 26]. Moreover, we considered several safety-only and security-only analysis frameworks, to see if they had been extended to handle safety-security combinations. In particular, we looked into threat analysis [13, 62]: however, to the best of our knowledge, no threat analysis framework tackles safety-security interactions in a model-based fashion. After gathering selected literature, publications were independently assessed by three reviewers following the methodology showcased in [57]. They were included or cited only upon agreement of two or more reviewers. Papers that were included in our final selection are listed in Table 1, together with number of citations taken from Google Scholar. The following paragraph showcases literature that was considered but then excluded from our final selection.

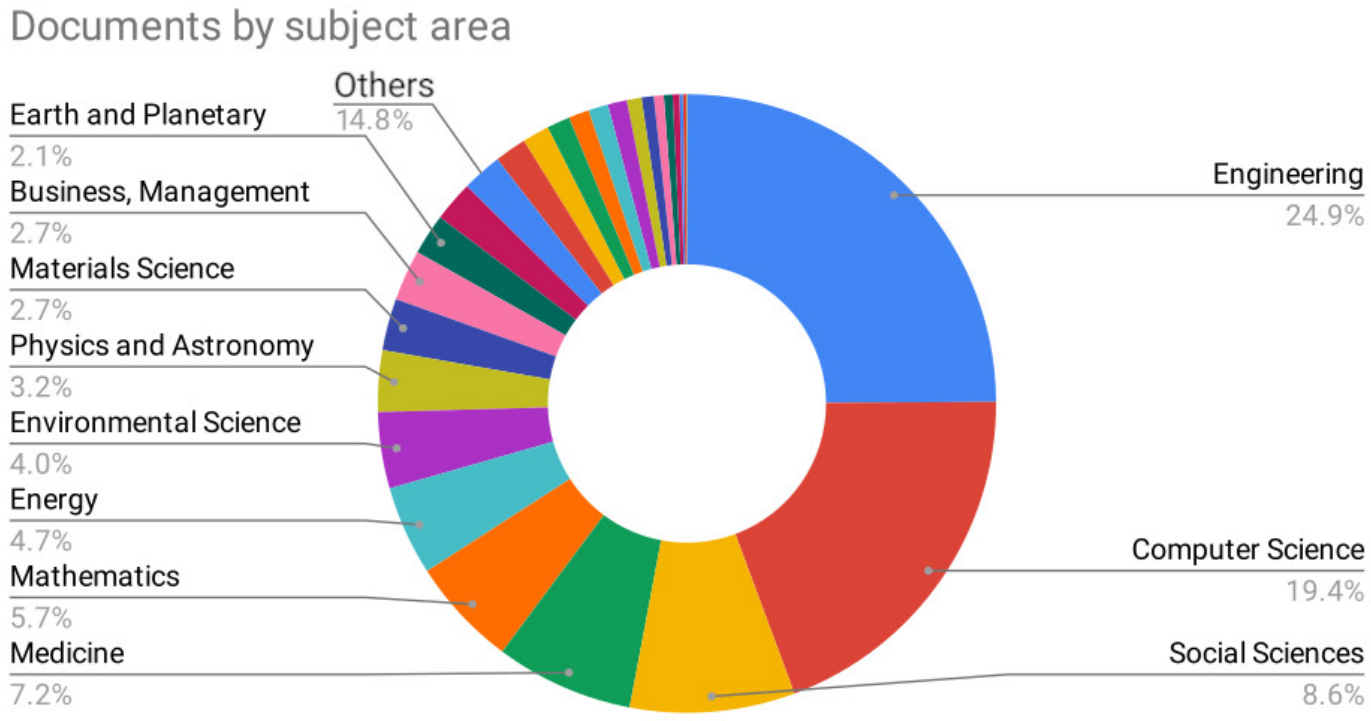

Figure 1: Percentages of published work that mention "safety" and "security" in title, abstract or keywords, divided by research field (time frame: 1922 - 2020). Data source: Scopus.

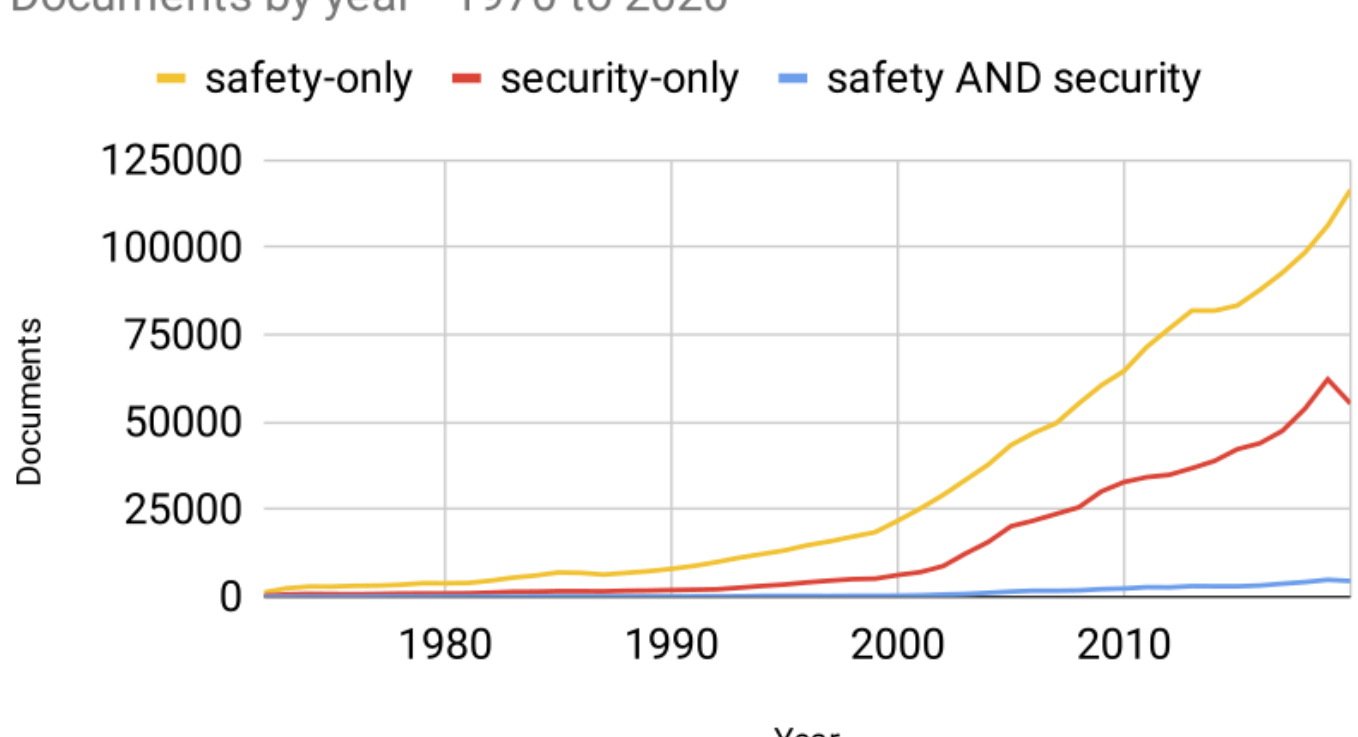

Figure 2: Number of published documents with "safety", "security" and both "safety" and "security" in title, abstract or keywords, ranging from 1970 to 2020. Data source: Scopus.

**Excluded papers.** During our literature review, we encountered papers on related topics regarding safety-security, which do not match the focus of this survey. We encountered a few papers on S-Cube [50, 52]. This modeling technique is very relevant for modeling safety and security interactions, however it seems to be focused on SCADA systems specifically. A paper on SOTERIA [25] presents modeling of IoT device networks for automated safety and security analysis from the device source code. This modeling relates specific to IoT devices. In this survey, we are interested in more general modeling techniques. There were several methods based on textual approaches, such as CHASSIS [79, 70], FMVEA/FMEA [79, 80, 86] and SAHARA [58]. However, they are based on a textual structure, while this survey focuses on model based approaches.

**Bibliometric analysis.** The areas of computer science and engineering cover about 44.4% of the published research on safety and security in the considered time frame, thus being the two major fields interested in this topic (as seen in Fig. 1).

Moreover, by comparing the bibliometric data regarding literature about safety, security and both safety and security we uncovered a growing interest in these fields since 1970s, testified by an increasingly higher number of publications mentioning these terms in the abstract, title or keywords. As shown in Fig. 2 the amount of publications that mention both safety and security in these fields is also increasing, while remaining at a considerably lower number when compared with safety-only and security-only literature. While the interest for their interactions is rising, the low amount of publications on safety and security suggest that there is still a considerable amount of work to be done in order to address challenges and open problems in the area of safety-security co-analysis (cfr. the blue line in Fig. 2).

## 3. Background on Attack Trees and Fault Trees

Since several safety-security formalism combine attack trees and fault trees, we briefly introduce these formalisms before discussing their combinations. In Section 4, we compare plain combinations of Fault Trees (FTs) and Attack Trees (ATs), while Section 5 surveys FTs and ATs combinations with additional features.

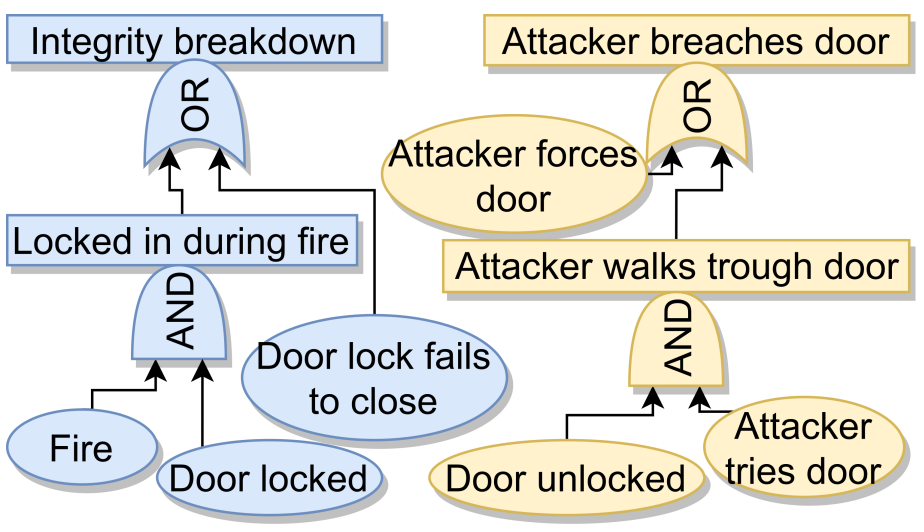

Figure 3: Fault Tree and Attack Tree

**Modelling.** FTs and ATs are hierarchical diagrams that model how low level failures (resp. attacks) propagate through the system and lead to system level failures (resp. attacks). Despite their name, FTs and ATs are Directed Acyclic Graphs (DAGs), rather than trees, since subtrees can have multiple parent gates. Furthermore, FTs are part of international standards [39]. FTs were developed in the early '60s [33] alongside Fault Tree Analysis (FTA) [76] and, due to their popularity, ATs were proposed in the '90 as their security counterparts [81]. FTs (resp. ATs) start with a Top Level Event (TLE), modeling a system level failure (attack), which is then refined through Boolean gates: the AND-gate indicates that all children must fail (be attacked) in order for the gate to fail (be attacked). For the OR-gate to fail (be attacked), at least one of its children need to fail (be attacked). When refining is no longer needed, one arrives at leaves of the tree: the Basic Events (BEs) in FTs model atomic failures; the Basic Attack Steps (BASs) in ATs model atomic attack steps. Non-leaves nodes (Fig. 3), such as *Locked in during fire* and *Attacker walks through door*, are called *intermediate events*, denoted by rectangles. The FT in Figure 3 models that integrity is compromised if one is either locked in during a fire or the door lock fails to close. This is modeled via the top OR-gate. One is locked in during a fire if there is a fire and the door is locked. This is modeled via the AND-gate connecting the BEs *Fire* and *Locked*.

**Analysis.** FT and AT enable different kinds of analyses [47]: *qualitative analyses* include Minimal Cut Sets (MCSs), indicating which combinations of BEs or BASs lead to the TLE. The set {Fire, Door locked} is a cut set in Figure 3. Quantitative analyses compute *dependability metrics*, such as the system reliability, attack probabilities and costs. For example, by equipping the BEs and BASs with probabilities, one can compute the likelihood of a system level failure or attack to occur. FTs and ATs have been used to analyse numerous case studies [92, 36, 23].

**Observations.** Apart from similarities, FTs and ATs also feature some remarkable differences: FTs often focus on probabilities, whereas ATs consider several other attributes, like cost, effort and required skills [21]. The difference with respect to the OR-gate between FTs and ATs is that in FTs, the probability of failure asks for total probability. Thus the (probability) value of an OR-gate is the sum of the values of its children, minus the value of their intersection. In ATs instead, attacks are characterised by their max probability. So the value of an OR-gate in ATs is the max value among its children. Furthermore, FTs have been extended with repairs [75], and dynamic gates [43, 30]; ATs with defenses, and sequential AND (SAND) gates [46, 36].

## 4. Category 1: formalisms combining Fault Trees and Attack Trees

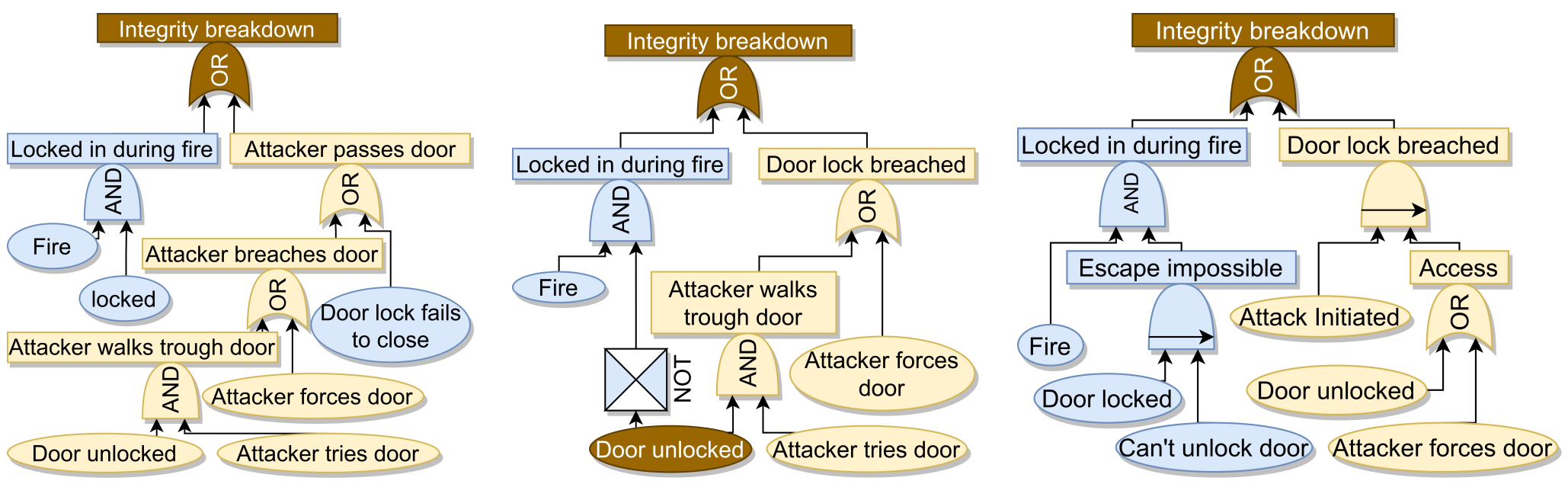

Figure 4: Lock door example as (a) FT/AT, (b) component fault tree and (c) attack-fault tree.

We first survey formalisms that combine FTs and ATs: *Fault Tree/Attack Trees* refine the basic events of a fault tree by an attack tree; *Component Fault Trees* merge fault trees and attack trees; and *Attack-Fault Trees* merge dynamic fault trees and dynamic attack trees. In all figures, we use blue for safety-events, yellow for security, and brown for their combinations. In the following paragraphs, we highlight answers to research questions with bold placeholders, e.g., (**A1**) denotes an answer to the first research question for the considered formalism.

### *4.1. Fault Tree-Attack Trees*

**Modelling.** Fault Tree/Attack Trees (FT/ATs) [35] are based on the assumption that attackers try to force a system failure by triggering the BEs in a FT. Thus, (**A2**) a FT/ATs indicates how the BEs can be attacked by refining these BEs by ATs with the BE as goal. By making this BE the goal of the AT depicted in yellow, Figure 4a details the basic event *Door lock fails* in Figure 3. (**A1**) FT/ATs do not provide a clear method for modeling dependencies of components in the AT on parts of the FT, so mutual reinforcement can not be modeled due to a lack of bidirectionality. Conditional dependency and antagonism can be modeled as per Section 6.

**Analysis.** The paper [35] (**A3**) computes the probability for the TLE to occur, given probabilities for the BEs and and the BASs.

**Observations.** Exploiting safety faults is a common method for hackers to enter a system, e.g.,: triggering a fire alarm to disengage a fire safety lock. Thus, it seems reasonable to consider the leaves of a FT as vulnerabilities, refining them via ATs. The fault tree's TLE then becomes the target for hackers. This can be reasonable e.g., shutting down a factory for ransomware. However, attackers are often exploiting safety faults to achieve other goals, such as stealing digital assets (e.g., Stuxnet worm targeting the availability of nuclear power plants to damage the nuclear program of Iran [45]). In that case, the forcing of BEs is only a starting point in an attack. (**A5**) Subsequent attack steps could be modeled in an AT; this could be a natural extension for FT/ATs. (**A4**) FT/ATs are used to model a case study on toxic chemical spill at a chemical plant in [35].

### *4.2. Component Fault Trees*

**Modelling.** Component Fault Trees (CFTs) equip FTs with a modular structure [44], so that a large FT can be modeled and analyzed in terms of smaller components. (**A2**) The paper [83] extends CFTs with security aspects by introducing a new BEs type for security breaches, which are essentially BASs. CFTs make no distinction between BEs and BASs. This is exemplified in Figure 4b, where attacks and failures are freely merged.

(**A1**) As per Section 6, CFTs model mutual reinforcement, conditional and antagonistic dependencies. In particular, antagonism is achieved by connecting one event A to an intermediate safety event B and an intermediate security event C, but having one of those connections be through a not gate (see Fig 5).

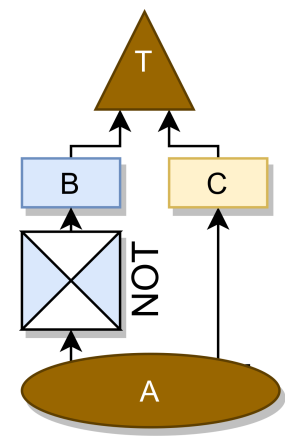

Figure 5: CFT antagonism.

**Analysis.** (**A3**) The CFTs from [83] enable the same analysis methods as FTs. The authors remark that, if a BAS only occurs in MCSs with multiple, low probability and independent BEs, then this attack is very unlikely to ever cause a disruption, as it requires multiple other unlikely events to occur. However, one may remark that the same holds for any event.

**Observations.** (**A5**) Not distinguishing between safety and security has the advantage that all existing tools remain applicable. Moreover, one may ask if it really matters whether a disruption is due to a failure or an attack. A disadvantage of merging failures and attacks, however, is that no interactions or trade offs between safety and security can be studied. (**A4**) Paper [83] models an Adaptive Cruise Control system as a case study.

### *4.3. Attack-Fault Trees*

**Modelling.** Attack-Fault Trees (AFTs) [54] treat attacks and failures in the same way CFTs do. However, (**A2**) AFTs merge *dynamic* ATs and *dynamic* FTs. These provide additional gates to model dynamic behaviour: dynamic ATs include the sequential AND-gate (SAND), modelling attacks as sequence of steps [9]. Dynamic FTs include gates for modeling spare components, functional dependencies (FDEP) and priority ANDs. Further, attacker profiles quantify the BASs with attacker's capabilities, such as resources, skills and damage. Figure 4c presents the Locked Door Example as an AFTs. The various SAND-gates indicate the order of events. E.g., for the *Door lock breached* event to happen, first *Attack initiated* must happen, and then the *Access* event as well.

(**A1**) AFTs can model conditional dependencies more explicitly via the FDEP gate: the trigger of the FDEP (i.e. leftmost input) automatically makes the dependent events (i.e. the other inputs) fail or be attacked. Even though AFTs do not include a NOT-gate, antagonism between events $A$ and $\neg A$ can be expressed through IFAIL nodes $A$ and $\neg A$, where $A$ has probability 0 if the $\neg A$ is activated, and vice versa (Fig 6).

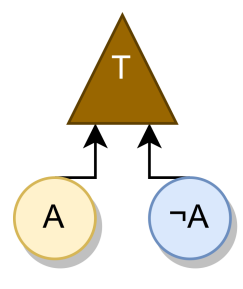

Figure 6: AFT antagonism.

**Analysis.** (**A3**) AFTs are analysed via statistical model checking by translating the AFTs to a network of stochastic timed automata. Using the attribute values in the attacker profiles, several metrics can be computed, such as the time, cost and likelihood of the attacks. Pareto frontiers elucidate trade offs between attributes, e.g., the likelihood of an attack within a given budget.

**Observations.** (**A5**) AFTs share the two disadvantages common to all FT approaches in that there is no distinction between safety or security failures once the TLE occurs, and the methods of modeling interactions may not clearly highlight the antagonistic dependencies. The attacker profiles in AFTs support a wide range of quantifiable parameters, enabling versatile analysis and trade offs. (**A4**) Remarkably, AFTs are used to model the medium-sized case study of an oil pipeline, in addition to modelling the Locked Door Example.

## 5. Category 2: formalisms extending Fault Trees and/or Attack Trees

This section surveys combinations of FTs and ATs with additional features: *State/Event Fault Trees* join a fault tree-like model with Petri nets, *FACT Graphs* join FTs and ATs with countermeasures and the ability to capture dynamic behaviours using triggers, *Boolean Driven Markov Processes* extend FTs and ATs with Petri nets and triggers, and *Attack Tree Bow-ties* combine Event Trees with an FT/AT-like model.

### *5.1. State/Event Fault Trees*

**Modelling.** (**A2**) State/Event Fault Treess (SEFTs) [73] join FTs and Petri nets[3], expressing that certain failures can only happen in certain states. Whereas the leaves in ATs and FTs model atomic events, SEFTs deploy Petri nets to accommodate state changes inside basic events. In Figure 7, the *Door* component can move between the states *Unlocked* and *Locked*. These state changes are triggered by events, depicted as black rectangles, that can be exponentially distributed, deterministic, or triggered by other events. Both states and events can be communicated via the gates of the tree, via in and out ports. In this way, Figure 7 expresses that a fire casualty can only happen if the door is locked. Figure 7 models antagonism: the door should be unlocked to escape from *Fire* but locked to prevent failure of the AND-gate *Enter unlocked door*. Following [73], we would model each subsequent attack step in the attacker component: first, the attacker tries to enter the door and, if it is not open, he/she tries to force it (the *Force door* step). However, to better represent antagonism wrt. the *Door* component, we decided to model the *Attacker tries door* step by embedding an AND-gate in the *House* component.

[3] [73] says state charts instead of Petri nets, but we did not see any state chart constructs, like hierarchical composition.

(**A1**) Besides antagonism, SEFTs support (directional) conditional dependency and mutual reinforcement, as per Section 6. This is achieved by a state machine that will always activate a safety or security state (Fig 8). (**A5**) Reworking SEFTs so that conditional dependency from safety to security can also be accounted could address a gap in the formalism and be considered a desirable extension.

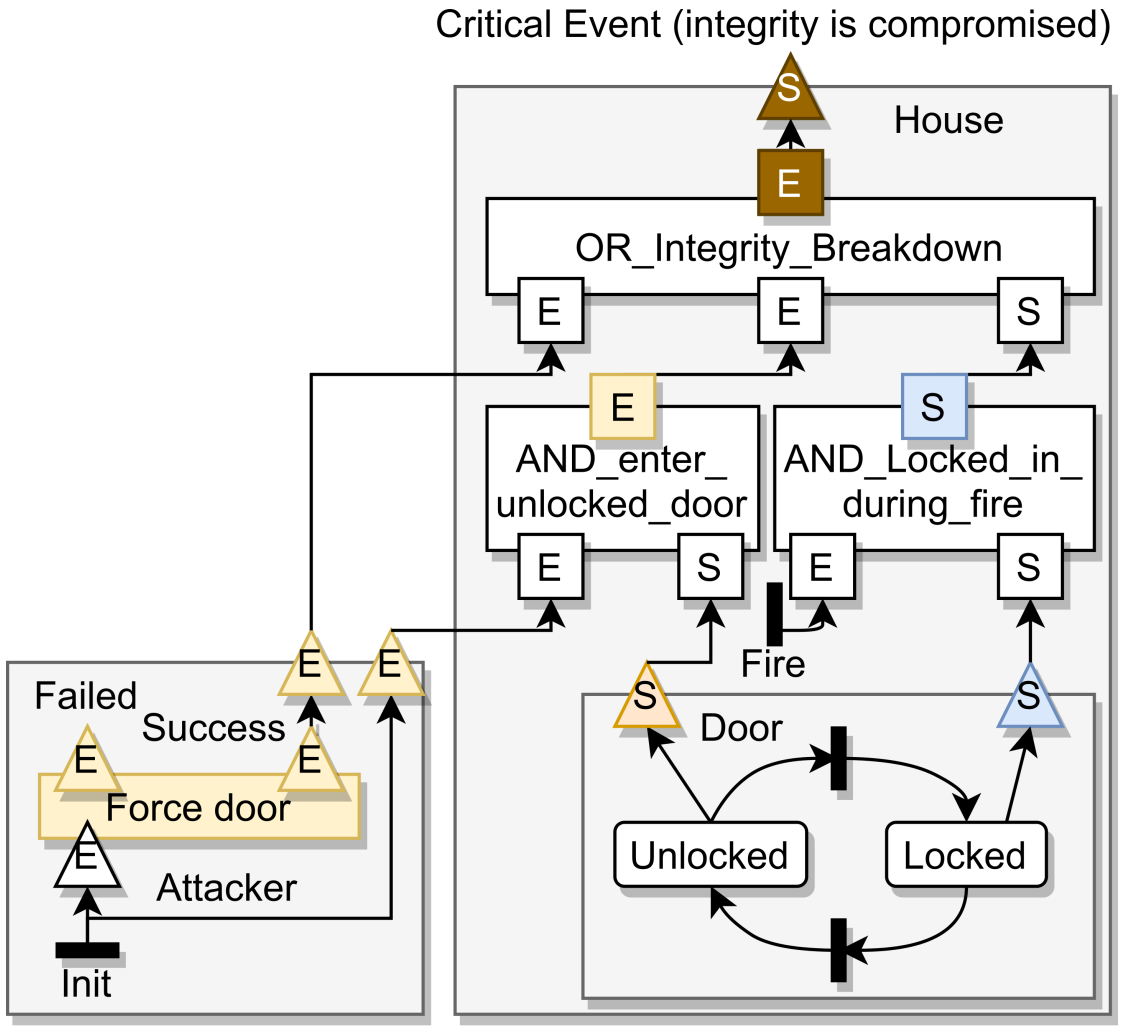

Figure 7: State-event fault trees.

**Analysis.** (**A3**) SEFTs support the same dependability metrics as attack-fault tree combinations. The tool ESSaReL translates SEFTs to extended deterministic stochastic Petri nets, which can be further analyzed by the TimeNet tool, e.g., for steady state analysis.

**Observations.** (**A4**) The authors of [73] model the small-sized example of a Tyre Pressure Monitoring System. Even though not mentioned explicitly in [73], it appears that Petri nets in the SEFT leaves must be disjoint between components. E.g., it is not possible to use the *Unlocked* state of the *Door* component as a direct input for an attack step. Allowing this could increase the expressivity of SEFTs.

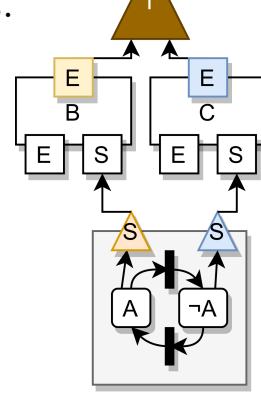

Figure 8: SEFT antagonism.

*5.2. Failure-Attack-CounTermeasure (FACT) Graphs*

**Modelling.** Failure-Attack-CounTermeasure (FACT) Graphs [77] (**A2**) join FTs and ATs with countermeasures and the ability to capture dynamic behaviours using triggers. To capture failures, attacks, and possible countermeasures the authors provide as a first step importing FTs at the end of the safety hazard and risk assessment phase. In the second step safety countermeasures are added to the graph. The third step adds an AT to the graph. Attacks are related to failures: much like FT/ATs, the elements of the FT are detailed with an OR-gate to specify that the failure can happen by itself or through the action of an attacker. In step four security countermeasures are added, possibly to any element of the AT. This results in a model similar to an FT/AT, with added countermeasures, as illustrated in Figure 9.

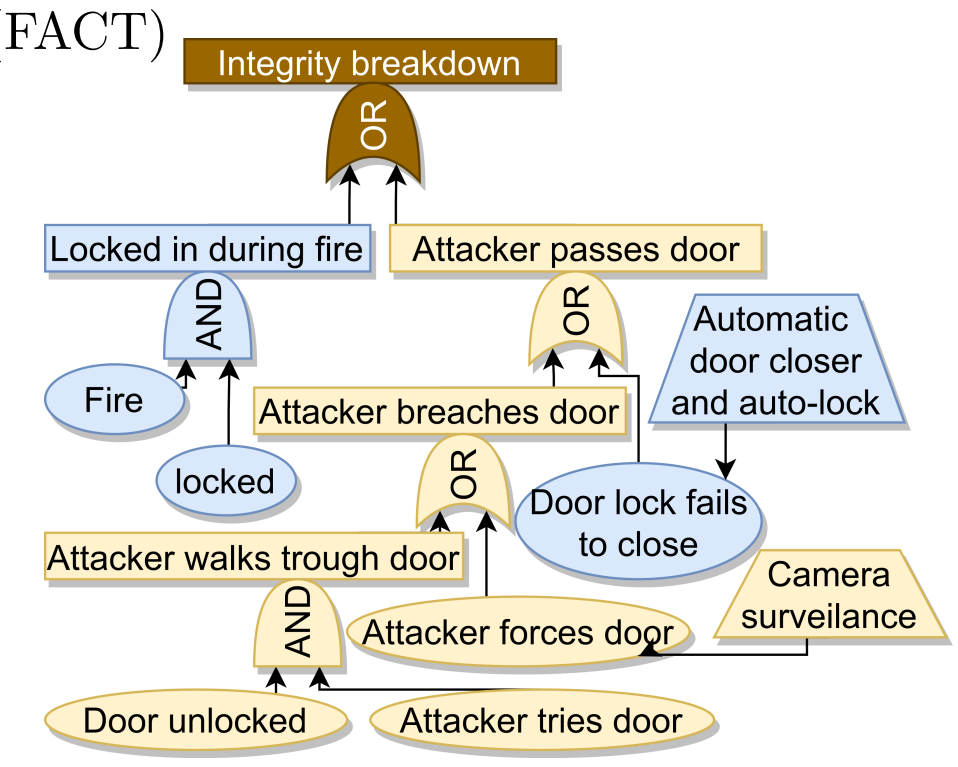

Figure 9: FACT Graphs.

**Analysis.** (**A3**) In spite of [77] not presenting any analysis of the proposed model, FACT graphs support analysis techniques like the ones enabled by FT/ATs. In addition to these, the presence of triggers and countermeasures could enable the analysis of the dynamic behaviour of the system and the role of countermeasures.

**Observations.** (**A1**) Similarly to FT/AT, FACT graphs can capture conditional dependency only directionally: in fact, FTs are detailed with ATs, but not vice-versa. (**A5**) Antagonism could be easily obtained in case a NOT-gate is added, filling a sensible gap. In [77], (**A4**) the authors focus on cyber-physical systems (CPS) and model vessel overpressure with a FACT graph.

### *5.3. Boolean Driven Markov Processes*

**Modelling.** (**A2**) Boolean Driven Markov Processes [15] extend FTs by equipping each leaf with a Markov process (MP), representing the different modes a component can be in. Various templates provide standard MPs to model standard failure behavior. For example, the *failure in operation* MP contains two modes, *operational* and *failed*. One transitions from *operational* to *failed* with an exponential failure rate $\lambda$, and back to operational with a repair rate $\mu$. The IFAIL MP models instantaneous failures. Moreover, users can define their MPs as a stochastic Petri net [51], similarly to SEFTs. In [14], Boolean driven Markov processes (BDMPs) are extended with security aspects by providing additional Markov processes for attacker steps. For example, the *Attacker Action* MP (AA) contains tree modes: in *Idle* the attacker has not yet initiated the attack. The *Active* mode corresponds to actual attempt, requiring an exponentially distributed time to succeed, and leading to the *success* mode. Further, *triggers*, represented by dotted red arrows, allow one MP to trigger a mode change in another MP. Figure 10 shows the Locked Door Example from [51]. The triggers pointing from *Attack initiated* to the OR-gate means that the OR-gate is not activated until *Attack initiated* happens. (**A1**) The use of Petri nets and the presence of intermediate events allow BDMPs to model all dependencies between safety and security, as detailed in Section 6. Antagonism is achieved by assigning the same Petri net to two leaves, but activating one node on one state, and the other when not in that state (Fig. 8).

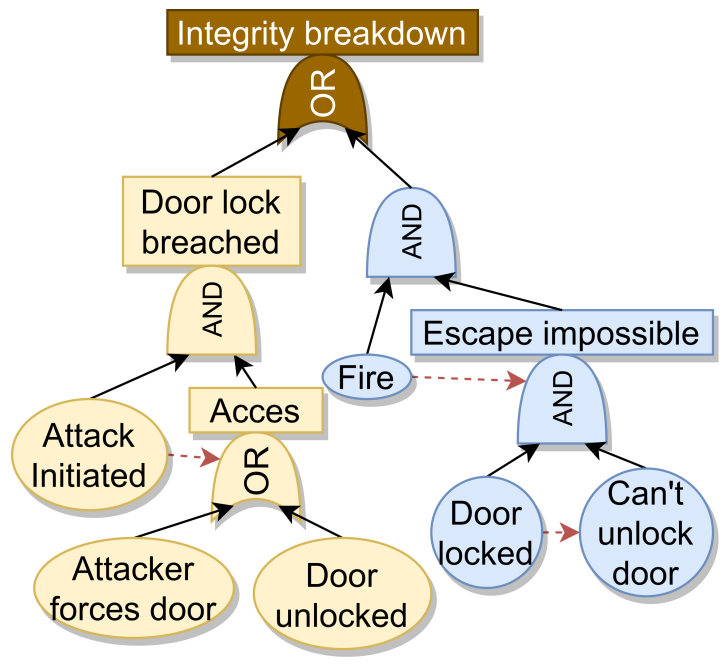

Figure 10: BDMP. Dashed lines in red denote triggers.

**Analysis.** (**A3**) BDMPs allows both quantitative and qualitative analysis. The authors of [14] build BDMPs with the KB3 modeling software platform. The computation of the overall mean time to success (MTTS), the probability of success in a given time and the list of possible attack success sequences (ordered by decreasing probability) is possible.

Figure 11: BDMP antagonism.

**Observations.** [22] is (**A4**) one of the few analyzed papers with a medium-sized case study, referring to an oil/gas pipeline. Furthermore, the authors discuss dependencies corresponding to those mentioned in the introduction. (**A5**) As an observation, in BD-MPs negation is not explicitly encoded in the structure but in underling Petri Nets for the leaves. An extension indicating such a negation similarly to how triggers are shown would be desirable. In Sec. 6 we propose a more general definition of dependencies in tree-like formalisms, thus dependencies definitions presented in [22] do not coincide with ours.

### *5.4. Attack Tree Bow-ties*

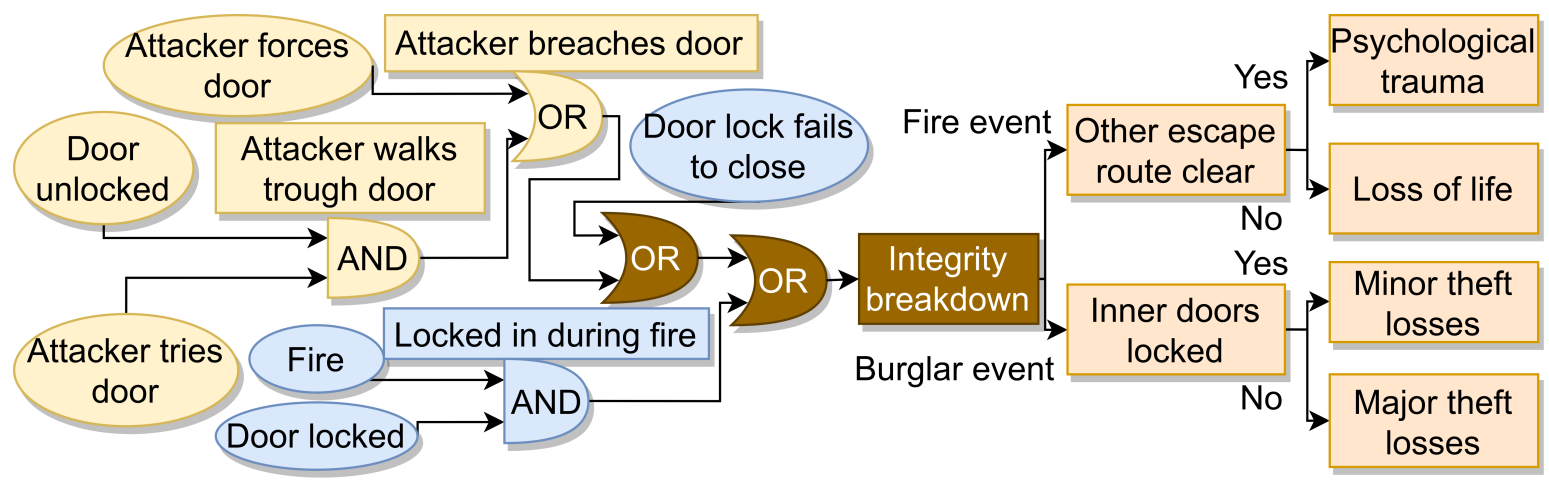

Figure 12: Attack tree bowties.

**Modelling.** (**A2**) Attack Tree Bow Ties (ATBTs) [3, 2] combine Bow-ties with ATs. Bow-ties [29] themselves combine FT and Event Trees (ETs): the left part of a Bow-tie is a FT modeling the causes of an hazardous event, which is in the middle of the Bow-tie. The ET on the right models its consequences. Barriers, i.e., a measure $M$ preventing some failure $F$ from happening, are modeled as $\text{AND}(F, M)$, so that the failure $F$ only propagates if the barrier $M$ fails. Now, ATBTs extend regular Bow-ties by attaching ATs to the basic events of a FT, just as in FT/ATs. Thus, in ATBTs, the left part of the Bow-tie is an FT/AT, rather than an FT. Figure 12 models the Locked Door Example via ATBTs. The leftmost part is equivalent to the FT/AT from Figure 4a. The ET on the right details the consequences in several cases: if the failure was a fire or a burglary, if alternate escape routes were available or inner doors were locked. (**A1**) Dependencies expressed in ATBTs are the same as in FT/ATs, as further explained in Section 6.

**Analysis.** (**A3**) Just like for regular Bow-ties, [3, 2] identify vulnerabilities via MCSs. By assigning likelihood level to all BEs and BASs, two likelihood levels are assigned to these cut sets: one for safety and one for security. Thus, trade offs can be made.

**Observations.** Since the left, causal part of ATBTs is similar to FT/ATs, similar observations apply. (**A5**) Furthermore, it would be natural to study trade offs between safety and security by equipping ATBTs with two hazardous events, one for safety and one for security. Moreover, ATBTs are one of the few formalisms that consider safety-security trade offs. (**A4**) In [3, 2] the authors create a small case study of a risk scenario in a chemical facility. In [8], the authors investigate the Facebook DNS outage from 2021 to show safety-security dependencies. They extend disruption trees which were introduced by [84] and add barriers (safety) and mitigations (security) in an attempt to model the four dependency types.

## 6. Dependencies in ATs and FTs Combinations

Below, we propose more precise definitions of the dependency types from [53] for the specific context of FTs and ATs. We focus on *events* in FTs, ATs and their combinations.

These events correspond to requirements: the TLE provides the main disruption to be avoided; the main requirement. The TLE refines into sub-events to be prevented; the sub-requirements. As such, these events and their interactions are the inverse of the logical interactions between the requirements. For each dependency type, we investigate to what extent these are expressible in the various FT and AT formalisms.

**Antagonism.** Two undesirable events $A, B$ are *antagonistic*, or *conflicting*, with respect to $C$ if at least one of them always occurs due to $C$. In tree-based formalisms, let $A$ and $B$ have a shared security/safety event $C$ as a child. Let $C$ be connected to either $A$ or $B$ through a NOT-gate. $C$ (not) occurring will either trigger $A$ or $B$ directly, or the other through the NOT-gate.

The events *Locked in during fire* and *Attacker walks through door* in Figure 4b are antagonistic with respect to *Door unlocked*. As such, expressing antagonism requires a form of negation. CFTs use NOT-gate to express negation. AFTs do so by tweaking (in a somewhat artificial way) the parameters of IFAIL. SEFTs and BDMPs express antagonism through a state-based model, where a system can be in only one state, e.g., door open or door closed. Like standard FTs and ATs, the FT/ATs and ATBTs models cannot model antagonism. However, a NOT-gate could easily be added.

**Conditional dependency.** If an undesirable event $B$ is conditionally dependent on event $A$, then event $B$ not occurring is only possible if event $A$ has not occurred: $A$ occurring implies $B$ occurring. In tree-based formalisms, if we make $B$ the TLE, and $A$ a leaf, each set containing $A$ must be a cut set for $B$. In Figure 4b, *Attacker forces door* is a condition for *Integrity breakdown*. All tree-based safety-security formalisms can express conditional dependencies between events. Since FT/ATs and ATBTs refine the leaves of a FT by an AT, they contain only paths from security to safety events. Thus, safety requirements can depend on security, but not the other way around. That also holds for SEFTs, where attack steps cannot depend on system's components that capture safety events. Further, the Functional Dependency (FDEP) gate in AFTs supports conditional dependencies, where $B$ occurs as soon as $A$ does.

**Mutual reinforcement.** Event $A$ reinforces event $B$ if the consequences of $B$ are less likely to happen due to event $A$. In tree-based formalisms, event $A$ reinforces event $B$, if every time $B$ appears in a cut set, $A$ does so as well. Events $A$ and $B$ mutually reinforce if the reverse also holds. Mutual enforcement typically occurs due to AND-gates, where both $A$ and $B$ are exclusively connected to the same AND-gate, either directly or through other (S)AND-gates. This configuration is expressible in CFTs, AFTs, SEFTs and BDMPs.

**Independence.** Two events $A$ and $B$ are statistically independent if $P[A \& B] = P[A] \cdot P[B]$. By assumption, all leaves in FTs and ATs are statistically independent. Events that are (mutually) reinforcing can also satisfy this statistical independence requirement. Thus, in tree-based formalisms the absence of (mutual) reinforcement is also required to capture Independence as intended. All tree-based formalisms from Sections 3 to 5 can express independence.

## 7. Category 3: Mathematical Formalisms

Two formalisms are based on modeling outcomes and interactions primarily mathematically, namely Bayesian Networks and an unnamed framework based on Calculating the impact of Threats, Hazards, and Opportunities, which we will call the THO framework. The THO Framework allows analysis of threats, hazards and opportunities of a given system and presents similarities with ATBTs.

### *7.1. Bayesian Networks*

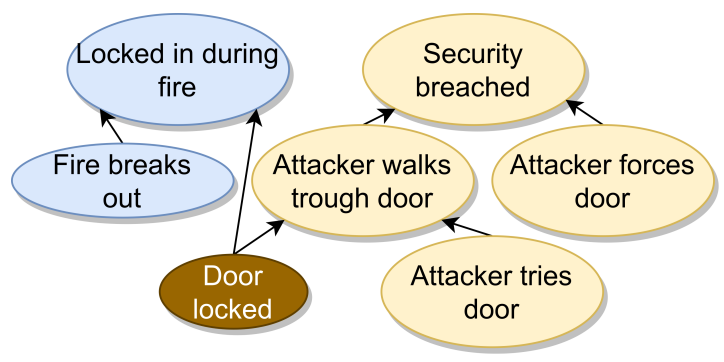

Figure 13: Bayesian networks.

**Modelling.** (**A2**) A Bayesian Network (BN) is a probabilistic graphical model that represent probabilistic dependencies between several variables via a directed acyclic graph. Each node $A$ represents a variable, and an edge from $A$ to $B$ indicates that A stochastically depends on $B$. A conditional probability table yields the conditional probabilities $P[A|B]$. If the probabilities of the leaves in the BN are known, the probabilities of the root nodes can be calculated. In [48], BNs are proposed to model safety and security dependencies. The two root nodes represent system safety and security. Figure 13 expresses that *Door locked* is a common factor for safety and security.(**A1**) BNs model conditional dependency, mutual reinforcement, and antagonism with a stochastic dependency in the DAG and appropriate values in the Conditional Probability Tables (CPTs). Antagonistic dependencies in particular are modeled by having a safety B and a security node C both depend on A, where the failure probability for one is higher when A is true, and for the other is higher when A is false: $P[B = 1|A = 1] > P[B = 1|A = 0] \wedge P[C = 1|A = 0] > P[C = 1|A = 1]$. (Fig 14)

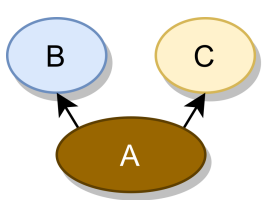

Figure 14: Bayesian network antagonism.

**Analysis.** (**A3**) Qualitative analysis can be done in the form of conditional independence analysis (analyzing which nodes influence other nodes and how). BNs enable quantitative analysis to calculate reliability metrics such as mean time to failure.

**Observations.** Fault trees and attack trees can be seen as special cases of Bayesian networks, where the CPT encode the Boolean gates, e.g., $P[A = 1|B = 1, C = 1] = 1$, 0 for other $B$ and $C$, for an AND-gate $A$ with children $B$ and $C$. Thus, the BNs extend ATs and FTs with flexible dependencies, and enable separate TLEs for safety and security. Furthermore, events after a TLE failure could potentially also be modeled. [10] discusses modeling Bowties with conditional probabilities, which appears to be a perfect example of such an extension. However, the gates in ATs and FTs provide a clearer visualization of the behavior, since in BNs these must be read from the probability tables. (**A4**) Paper [48] uses the pipeline example as a case study. (**A5**) BNs can also be extended to Dynamic Bayesian Networks (DBNs), where multiple copies of the BN represent the state at different time steps. Nodes in the DBN can depend on nodes in previous time steps. More complex gates like sequence enforcers can then be modeled [63].

### *7.2. Threads-Hazards-Opportunities Framework*

**Modelling.** [10] presents a framework to perform risk analysis for both safety and security. When modeling a given system, the framework considers possible threats and hazards that can lead to various consequences, or *outcomes*. Opportunities — such as e.g., a planned shutdown, which allows for preventive maintenance — are also considered. Consequences would be typically expressed by real values representing *observable quantities* for e.g., economic loss, number of fatalities, number of attacks, the proportion of attacks being successful. The author provides steps that describe the risk and vulnerability analysis process: 1. Identify the relevant functions and subfunctions to be analysed, and relevant performance measures (observable quantities). 2. Define the systems to meet these functions. 3. Identify relevant sources (threats, hazards, opportunities). 4. Perform an uncertainty analysis of the sources 5. Perform a consequence analysis, addressing uncertainties. 6. Describe risks and vulnerabilities. 7. Evaluate risks and vulnerabilities. 8. Identify possible measures, and return to 3. After identifying the relevant functions and subfunctions to be analysed (Step 1), the system in question is defined (Step 2): understanding how the system works is a key step, so departures from normal, successful operation can be easily identified. In Step 3 threats, hazards and opportunities are identified. This can be done e.g., through analysis of statistics or through tools such as FMEA/FMECA/FMVEA [79, 80, 86] and HAZOP [31]. This step is heavily integrated with Step 4, performing an uncertainty analysis of sources. Once initiating events are identified and attackers' resources are assessed, event trees are used to develop scenarios starting from the initiating events. (**A2**) Once specific scenarios are identified e.g., a burglar entering the house, standard analysis can be performed using event trees and FTs (Step 5). In Step 6 risks and vulnerabilities are described. Specific quantities are selected e.g., the number of future attacks, the proportion of the attacks being successful, the number of successful attacks, and then uncertainties are assessed using probabilities: this leads to probability distributions of the above quantities (details on this in the following paragraph). Finally, Step 7 and 8 take place: however, [10] does not provide further details on these.

**Analysis.** (**A3**) For the quantitative analysis the uncertainties are expressed with probabilities and expected values $E(X)$ for the uncertainty distribution of $X$. For example $X$ can take one of the values 0, 1, 50 and the associated probabilities are 0.8, 0.11 and 0.05, then the expected value is $E(X) = 0 \cdot 0.8 + 1 \cdot 0.11 + 50 \cdot 0.05$. In this framework, probability is used as a measure of uncertainty, seen through the eyes of the assessor. In contrast to BNs, when [10] uses the notation $P(C|D)$, it does not indicate the probabilities of $C$ depend on the result of $D$. The probability assignments are dependent on available information and knowledge of the system. $P(C|D)$ indicates the Expected value of $C$ depends on $D$ with unchanged probabilities. For example with sufficient information we are able to predict with certainty the value of the quantities of interest. The quantities are unknown to us as we have lack of knowledge, i.e., how people act, how machines work, etc. With the help of this framework one can calculate the likelihood of an attack and its consequences to failures.

**Observations.** (**A4**) In [10], the authors utilize the THO Framework to model the case study of an electrical grid. (**A1**) Given that [10] proposes to utilize both FTs and event trees, and to assess the impact of vulnerabilities/attacks on a given system, a clear parallel with ATBTs can be drawn. FTs can be employed to assess hazards on a given system, and we could evaluate the role of attacks by using ATs as seen in e.g., FT/ATs. Moreover, for every successful attack, consequences can be evaluated through the use of event trees. This structure suggests strong similarities with ATBTs. Thus, the framework proposed in [10] seems to capture dependencies in the same ways as ATBTs do (this also applies to our running example). (**A5**) It is however unclear if developing an ATBT for every identified scenario would be the best choice to represent the aforementioned procedure. Other options would include the development of an ATBT with multiple TLEs in order to represent diverse scenarios and their consequences.

## 8. Category 4: Architectural Formalisms

Several of the formalisms are based on modeling the system architecture and then verifying its correctness, namely STAMP, SysML, Alloy, Event-B, and AADL. These models primarily describe the overall structure, architecture, behavior or interactions of the system they model, and then check if the modeled properties comply to a set of requirements and/or conditions.

### *8.1. STAMP*

**Modelling.** The System-Theoretic Accident Model and Processes (STAMP) is rooted in the observation that system risks do not come from component failures, but from inadequate control or enforcement of safety-related constraints. Rather, (**A2**) in STAMP, systems are viewed as interrelated components that are kept in a state of dynamic equilibrium by feedback loops of information and control [56]. Each component enforces the safety and security constraints in the processes it controls, using control actions and feedback messages. Inability to enforce these constraints results in failures in safety or security. System Theoretic Process Analysis (STPA) and its extensions STPA-sec [78] and STPA-safesec [38] systematically identify the consequences of incorrect control actions and feedback [69], e.g., when these happen too early, in the wrong order, or were maliciously inserted. Figure 15 shows the *Person*, who can lock and unlock the door. *Locking mechanism* is controlled by *Person*, and controls if the *Door* can open. A safety constraint is that *Person* must be able to unlock a door in case of fire; a security constraint is an unauthorised person must not be able to gain access. A *violation* is, e.g., the scenario where the person locks the locking mechanism while the door is open, forcing the door to stay open and granting unauthorised access. (**A5**) STPA-safesec can discover these risks in a structured manner, however this is currently still a manual process requiring domain knowledge and has not been automated [68]. (**A1**) STAMP is not geared towards expressing dependencies. However, STPA-safesec analysis may reveal safety-security conflicts [91].

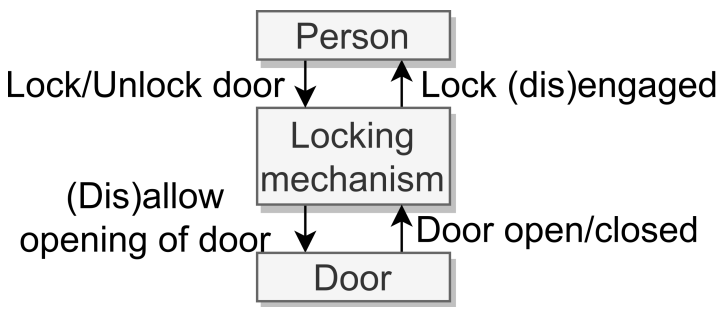

Figure 15: STAMP.

**Analysis.** (**A3**) STPA is used to identify potential hazards and undesired behaviours: e.g., the violation previously described. It would for example invert the order of control actions, such as locking the lock before closing the door, and identifying if that leads to a prohibited state. A door locked permanently open would violate security requirements.

**Observations.** STAMP models control flows of the system and further analysis identify safety and security hazards in that control flow. Domain experts are required to properly identify issues. STAMP provides a structured way of reasoning based on a high level description of the system that should ensure the identification of safety and security requirements. STAMP can identify safety and security issues in a system, it is not for documenting those interactions. It is suited to help with the syntheses of a model describing safety and security interactions. (**A4**) Paper [38] analyse synchronous-islanded operating microgrids using STAMP.

### *8.2. SysML*

**Modelling.** The Systems Modeling Language (SysML) is a general-purpose modeling framework for systems engineering, extending the Unified Modeling Language (UML). (**A2**) SysML-sec [74] extends SysML with safety and security requirements. Its functional model describes the communication channels between processes, detailing their encryption methods and the complexity overhead cost. SysML-sec also includes a system mapping model, describing which part of the communication process occurs in which components. SysML-sec enables safety and security properties to be expressed and verified via separate model checkers, e.g., checking if confidential communications can be intercepted by compromising a bus. The interaction between safety and security properties can, however, not be modeled. (**A5**) Since SysML-sec is geared towards embedded software, we replaced the Locked Door Example with a digital keypad lock. Figure 16 shows the mapping model. Within the keypad, the CPU is connected to the main memory and an Input-Output Bus. Connected to this bus is the physical button pad, as well as a Digital Analog Converter that engages and disengages the lock. Figure 17 details the process of (dis)engaging the lock. The correct codes are stored in an encrypted format, and an input code is received unencrypted. Both codes are collected, and the input code is checked against the stored key, either by decrypting the stored key or encrypting the input key. On a match, a command is sent to (un)lock the door.

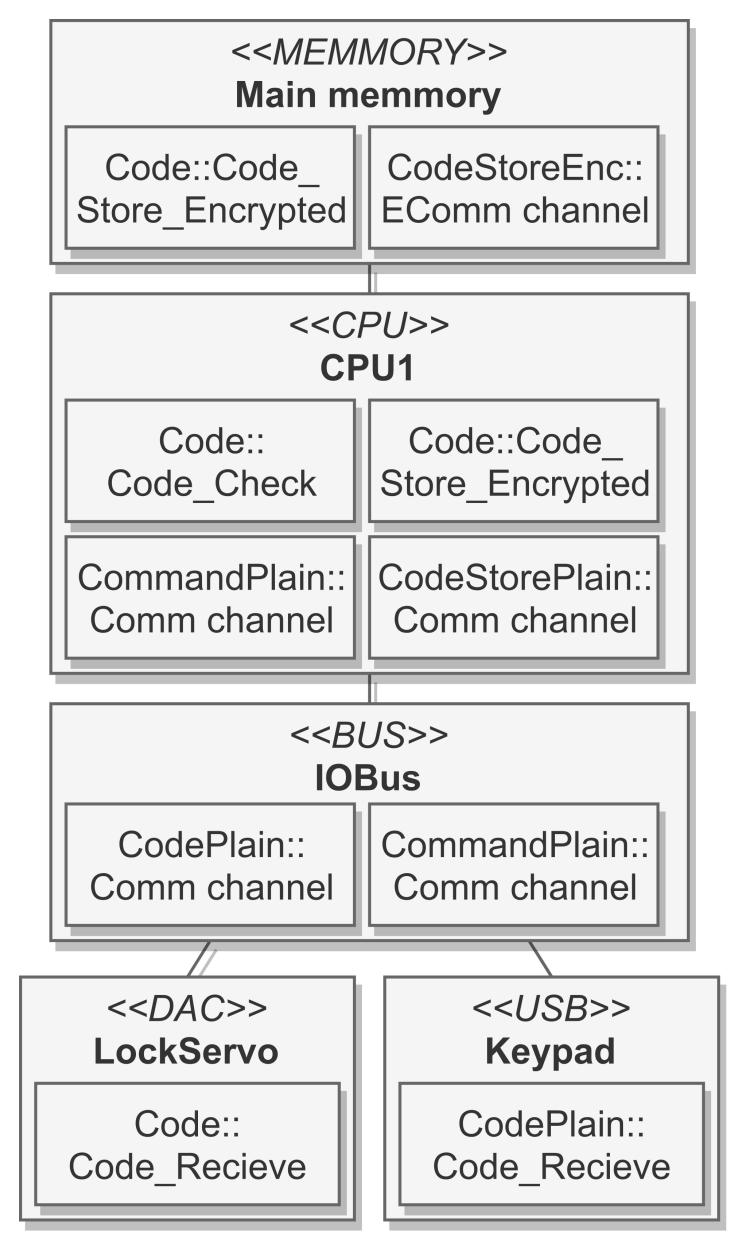

Figure 16: SysML mapping.

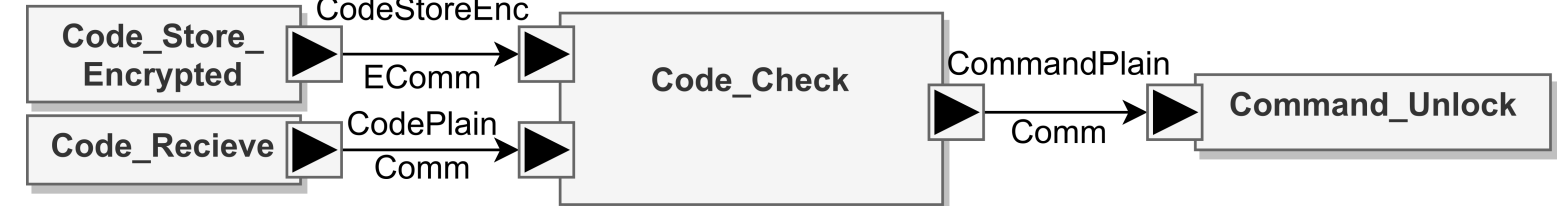

Figure 17: SysML function diagram for digital door

**Analysis.** (**A3**) The TTool [67] can encode ATs and FTs in a SysML parametric model. It then uses UPPAAL to test for reachability of the TLE. Counter measures can be annotated to the ATs and FTs models, and individual attack vectors/BEs switched on or off. UPPAAL will then indicate if preventing the chosen subset of attack vectors/BEs is sufficient to prevent the TLE. No method for discovering the ATs and FTs from data is included. TTool can also verify security requirements with ProVerif [6]. Requirements can be tagged as e.g. *Confidentiality*, *Non Repudiation*, *Data Origin Authenticity* [7]. A confidentiality requirement on all communications can be created, and communication channels from the functional model are then linked to these requirements. An unsecured bus does not satisfy the confidential requirement, and with no other mitigation, e.g. encrypted messaging, ProVerif will find a state where the confidential requirement of the communication channel is violated, returning a trace showing how this state was reached. Paper [67] uses Key Masters Keying Protocol, which aims to securely distribute a randomly generated key among a group of in-car Electronic Control Units, as a case study.

**Observations.** (**A1**) In SysML, safety and security are modeled and analyzed separately. Violation of safety and security requirements can be observed, as well as mitigation measures that violate other requirements. (**A5**) SysML does not provide tools for analyzing interactions between requirements, as interactions between safety and security are an emergent behavior of the models.

### *8.3. ALLOY*

**Modelling.** Alloy [42] is an Object Oriented system modeling language, using set theory to prove assertions on a given model. It consists mainly of *signatures*, which define the structure of classes of objects, *facts*, which define rules the overall system follows, and *assertions*, behaviors that the overall system should comply to, but counterexamples that violate them could exist for. The authors in [20, 19] propose (**A2**) Alloy as a basis for creating a system model, and then creating two derivative models, which separately check the safety and security requirements through assertions. This is similar to SysML, which also defines a base behavior of the system for which safety and security are checked separately. (**A4**) The literature we examined contained a case study for a fire safety system [19], but it did not contain a security element. Due to a lack of an industrial case study that fits our focus of safety-security co-analysis, we have crafted an Alloy model that captures the locked door example. The first step in defining the model was creating the mechanism for having a Boolean data type. This is achieved by extending an abstract Boolean into a singular *TRUE* and *FALSE*. In stead of requiring a new object for every attribute that could be added to a class, the name of the attribute in combination with pointing to a globally unique *TRUE* and *FALSE* would describe if the property is present or not. Next, *Doors*, *Locks* and the *Building* are defined. A door can be opened or not and a *Lock* engaged or not, state registered by a *Boolean*. A *Door* has a single *Lock* and a *Lock* can be installed in a single *Door*. There is also a *Building* that contains all the *Door*s, which can be on fire described by a *Boolean*. Next we impose some rules on the model to ensure accurate behavior by defining *fact*s. If a *Door* has a *Lock*, then that *Lock* must reflect it is installed in that *Door*, and if a *Lock* identifies a *Door* it is installed in, that *Door* must reflect it has that specific *Lock*. Lastly, all *Door*s that exist must be installed in the *Building* to exclude free floating *Doors*. Three *predicates* are defined, which can be used for evaluation. A *predicate* defining if a specific *Door* d is openable, a *predicate* defining if a person can always escape the *Building* fire, and a *predicate* defining if a burglar is prevented from entering. Lastly, *asserts* are defined, which are used to check if the above defined model fulfills the safety and security requirements. The current model does not fulfill the *asserts*, since the model does not prevent states that violate the *asserts*. However, introducing new *facts* to enforce the safety *assert* either violates the security *assert*, no doors are not openable results in all doors are openable, or requires the *Building* to never be on fire, which is unrealistic.

```
abstract sig Boolean { }
one sig TRUE extends Boolean { }
one sig FALSE extends Boolean{ }
sig Lock {lock_engaged: one Boolean, installed_in: one Door}
sig Door {opened: one Boolean, installed_lock: one Lock}
one sig Building {doors: set Door, on_fire: one Boolean}
```

```
fact {all d:Door, o: d.installed_lock | o.installed_in = d}
fact {all l:Lock, o: l.installed_in | o.installed_lock = l}
fact {all d:Door, b:Building | d in b.doors}

pred door_openable [d: Door] {d.opened = TRUE or
d.installed_lock.lock_engaged = FALSE}
pred can_escape_fire {no d: Door, b:Building |
b.on_fire = TRUE and not door_openable[d]
pred burglar_cant_enter {no d:Door | door_openable[d]}

assert fire_safe {can_escape_fire}
assert secure {burglar_cant_enter}
assert safe_and_secure {can_escape_fire
and burglar_cant_enter}
```

**Analysis.** (**A3**) Alloy is both a language and a toolkit: as such, Alloy models can be automatically analyzed. This is achieved by brute force SAT solving. Alloy can perform checks for a given amount of objects, either a specific amount of objects or all integers up to a specific amount. For these objects, Alloy checks if the assertions hold in all possible configurations of those objects, returning a counterexample where an assertion is violated. A graphical representation of such a counterexample can be automatically generated.

**Observations.** (**A1**) Alloy can model the absence of interactions by default. Due to the predicate logic, conditional dependency, mutual reinforcement, and antagonism can be encoded directly. Conditional dependency is achieved by defining a mainly safety or security focused predicate that requires a predicate of the other kind to be true. Mutual reinforcement can be modeled by having a safety and a security predicate requiring the same state of another predicate or of a component in the model. Antagonism can be directly modeled by defining a combined safety and security predicate or assert that simultaneously requires a predicate to be true and not true. For example, defining that all doors must simultaneously satisfy and not satisfy the door_openable predicate. Similar to SysML, antagonism can also be detected as an emergent property of the model, where changing the base behavior to satisfy a safety *assert* would violate a security *assert* and vice versa. However we do not count this behavior as explicitly modeling antagonism. While the Alloy language is capable of handling these interactions, using alloy as described in [42], that is, making separate adaptations of a base model to check safety and security separately, would preclude defining all these interactions. (**A5**) At the time of writing this survey, a new version of Alloy has just released adding support for some temporal relations. No literature exists yet that explores the possibilities of this extension in the context of safety and security. As such, we will consider only the older version of Alloy, and exclude the temporal element in our analysis for now.

*8.4. Event-B*

**Modelling.** Event-B [4, 5, 24] has been developed out of the framework RODIN [82]. It is a rigorous approach to correct-by-construction system development. Development starts from the definition of an abstract specification — which models the essential functionalities of the system — that is later refined. In the refinement process, the abstract model is transformed into a detailed specification, expressed using set theory and propositional logic. This propositional logic also supports expressing properties at differing time steps, for example, statements that the difference of variables over time is constrained ($|A(T+1)| < |A(T)| + |\Delta|$). (**A4**) In [89], the authors model the architecture of a battery charging system representing its failure behaviour and defining the mechanisms for error detection and recovery. (**A2**) During the refinement process, they also represent the effect of security vulnerabilities such as tampering, spoofing and denial-of-service attacks and analyse their impact on system safety. The step-wise refinement process of Event-B allows one to systematically derive the constraints and define the assumptions that should be fulfilled to guarantee system safety in presence of security attacks. This example is representative of conditional dependency, specifically of security potentially influencing safety. First, we detail the system by defining requirements and constraints. There are two requirements that encode the integrity of the system: 1. You must not be locked in during fire. 2. Attacker must not be able to open the door. There is a single constraint: 1. Door cannot be locked or unlocked in the same time. We assume that door opened = 1 and lock locked = 1, and formalize the state of the system using the variables $d$ and $l$. We then proceed to refine the model. First refinement: we add guards to the lock and the door events. Second refinement: we add a variable for fire, an invariant and event for fire safety: when fire happens the door must open, the lock must open to ensure the door can be opened. Third refinement: we add a variable for burglary and an invariant for burglary, plus an event for burglar security. The process of refinement unearths the need for the following property: $f = 0 \vee b = 0$. However, it is not realistic that fire and burglary will never happen simultaneously.

**Constants:**

**Variables:** $d, l, f, b$

**Invariants:**

Inv01: $d \in \{0, 1\}$

Inv02: $l \in \{0, 1\}$

Inv03: $f \in \{0, 1\}$

Inv04: $b \in \{0, 1\}$

Inv05: $f = 1 \implies d = 1$

Inv06: $b = 1 \implies d = 0$

**Properties:**

Prop01: $f = 0 \vee b = 0$

**Events:**

**OP-DOOR:** **when** $d = 0; l = 0$ **then** $d := 1$

**CL-DOOR:** **when** $d = 1; l = 0$ **then** $d := 0$

**OP-LOCK:** **when** $l = 1$ **then** $l := 0$

$$\textbf{CL-LOCK: when } l = 0 \textbf{ then } l := 1$$
$$\textbf{FIRE-SAFE: when } f = 1 \textbf{ then } l := 0; d := 1$$
$$\textbf{BURGLAR-SEC: when } b = 1 \textbf{ then } l := 1; d := 0$$

**Analysis.** (**A3**) Given some events for the system and its invariants which are the requirements of the system, Event-B can prove properties of that system (e.g., safety-related properties) under given assumptions [87]. However, note that security mechanisms and requirements are still modelled with a similar method as safety ones are, with the same limitations. (**A5**) Event-B is a text-based formalism, thus not allowing graphical visualization of modelled systems.

**Observations.** (**A1**) As shown, antagonism can be modelled in the lock-door example. Because Event-B is a very general approach - mainly based on set theory and propositional logic - it can model all four kinds of safety and security interactions. In particular, we can encode conditional dependency inside events using guards.

### *8.5. The Architectural Analysis and Design Language*

**Modelling.** The Architectural Analysis and Design Language (AADL) is a framework to model the software and hardware architecture of embedded real-time systems. It is an international standard of the SAE, Aerospace Division [1]. Its core language describes the multi-threaded, distributed software architecture, and the annexes describe real-time behaviour and error modelling. The EMV2 language is used to analyse software failures (safety), providing a property set, a library of error models, and an annex sub-language, expressing how errors are generated and propagated. (**A2**) Safety and security are added in one AADL model in [28] as follows: Safety analysis computes the mean time between failures (MTBF) of the system based on the information provided by the AADL model and its error annex. For security, the goal is to avoid unauthorized access to sensitive data. Data is accessed through AADL ports of subprograms. So it is important to verify that the indirect data access points are secure enough. This is done with a dedicated property set that associates a security level (an integer value) to AADL data. Figure 18 models the Locked Door Example with AADL: this is uncommon for AADL because it is often used for the interaction between software and hardware components. Here we consider a *Person*, a *Locking mechanism*, and *Door access* as components. They are linked through ports. (**A1**) AADL does not model any kind of dependencies: safety and security are considered separately, although they share the same AADL model to perform the analysis.

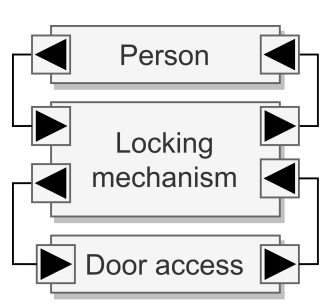

Figure 18: AADL.

**Analysis.** (**A3**) For safety, the MTBFs of the system is calculated via the AADL Error model statements in various components descriptions. They are compiled together to generate a FT. For security, the goal is to hide data with help of the Stood for AADL tool.

**Observations.** An AADL model is scalable and appropriate for version and configuration management, because it can be fully described by its textual representation. AADL can be also integrated into the MILS architectural approach that consists of developing an abstract architecture intended to achieve the stated purpose, and implementing that architecture on a robust technology platform [27]. FTs can be analyzed after transforming AADL to the Arbre Analyste tool, which makes AADL an excellent application for FTs. (**A5**) Interesting for future work is to integrate ATs with AADL. (**A4**) In [28], a toy example is presented, analyzing the safety and security of an electrical locked door. (**A1**) An AADL model does not represent any kind of safety-security dependency, thus it could be valuable to study whether the analysis for classical ATs and FTs combination could help understand if dependencies can be represented via safety-security measures.

## 9. Expressiveness

**Context of comparison.** In light of our analysis, we compare the formalisms with respect to their expressiveness, see Fig. 19. We do not consider AADL, SysML or STAMP in terms of expressiveness. While they are powerful system modeling techniques, they do not directly model the safety/security interactions of the system, which is the primary focus of this survey. Only AFTs, BDMPs, BNs, ALLOY and Event-B — although with different methods — fully capture all four interactions: antagonism, mutual reinforcement, conditional dependency and independence. The modelling techniques under analysis are vastly different and a one-to-one comparison would not be fair or complete. However, we are not trying to determine an objectively best modelling method. However, despite the difference of considered formalisms — especially when comparing architectural ones and formalisms combining/extending FTs and ATs — we believe a situational comparison can be drawn by focusing on the ability of each formalism to capture safety-security interactions, which is one of the key questions proposed in this paper. Moreover, all the observations regarding key similarities and differences highlighted in each respective section remain valid. Furthermore, we detail the expressiveness analysis of formalisms by considering the following properties: *ability to capture safety-security dependency relations*, *time*, *state of components*, *consequences* (see Table 2).

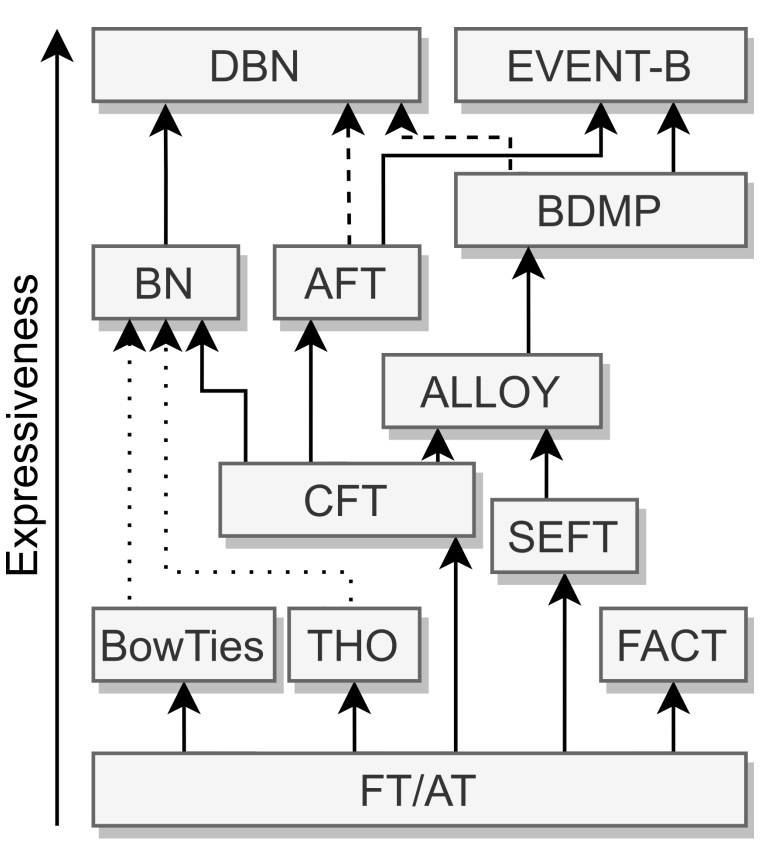

Figure 19: Expressiveness of formalisms: higher = more expressive. Solid arrows: technique subsumes in features. - - if DBNs support general probability distributions. ··· if BNs are expanded to model consequences beyond safety/security TLEs.

**Expressiveness.** FT/AT express independence, directional antagonism, and directional conditional dependency (from security to safety). ATBTs express the same plus consequences after failure. the THO framework mirrors ATBTs to a high degree, but is focused on expected values rather than probabilities. CFT express the same as FT/AT plus bi-directional conditional dependency/antagonism and mutual reinforcement. FACT graphs can model everything that FT/ATs can, but includes the ability to add countermeasures to security or safety risks. SEFTs express the same as FT/AT plus details on the state of components in the system via Petri nets. This enables negation of single components by depending on the non-failed state (an implicit negation), but does not enable negating sub-trees. Thus, SEFTs are slightly less expressive than CFTs (when NOT-gates are included). The attacker steps can also not be influenced by anything in the safety portion, so conditional dependency is directional too. ALLOY's predicate logic can model bidirectional antagonism and negation for entire sub-trees, like CFTs, and does so by exploring the possible states of components, thus is able to detail state like SEFTs, though purely for a qualitative analysis, since it lacks quantitative analysis. The current literature is also not up to date with the state of the ALLOY tool, which now includes the capability to model temporal relationships. When the capabilities for such are properly explored, it might be as expressive as EVENT-B. AFT express the same as CFT, plus an element of time/order through dynamic gates. AFTs use exponential probability distributions, as well as more complex distributions [54]. BNs model all dependencies, multiple TLEs, and potentially can be expanded to model consequences in a method similar to ATBTs, though they have not been used for such in the literature yet. Thus, the introduction of the dotted line in Figure 19. BDMP express bidirectional dependencies and details on the state of components via Petri nets, like ALLOY, plus time/order through the use of triggers. Recall they also express probability distributions as mentioned in Section 5.3. Dynamic BN express time, consequences, negation and details on the state of components. They also expand on AFTs and BDMPs if they model generic probability distributions. Event-B, due to its general nature, is the most expressive when it comes to qualitative analysis, though it lacks quantitative analyses. It supports state modeling, and can define system properties over time. Though not shown in literature, it could model consequences such as defined by ATBTs as part of system behavior.

## 10. Findings and Reflection

Our survey revealed several noteworthy findings, some of which are summarized in Table 2, Table 3 and Table 4.

**Finding #1: The majority of approaches combine Fault Trees and Attack Trees.** Seven of the fourteen formalisms we found combined FTs and ATs. This may not be too much of a surprise, since FTs and ATs are similar in nature and are widely employed, both in industry and academia. Some of these formalisms are plain combinations of FTs and ATs, while some others add constructs to extend modelling/analysis capabilities of these combinations.

**Finding #2: No novel modeling constructs are introduced.** It is however remarkable that none of the formalisms for safety and security integration that we found introduces novel modeling constructs to capture safety-security interactions. In our opinion, the reason for this phenomenon is to be attributed to familiarity with either safety-specific or security-specific formalisms. In fact, the key strategy that we unearthed is to *merge* already existing safety and security formalisms without adding new operators. Thus, one can represent safety and security features in one model, but one may wonder to what extent the interaction can be expressed appropriately when the main strategy relies on merging existing techniques.

**Finding #3: Safety-security interactions are still ill-understood.** While the paper [53] coins the four dependency types, these are given in natural language. As such, when one find him/herself trying to apply these definitions to capture interactions via formal methods, it remains unclear how one might translate them formally. In this work, we make a first attempt at proposing rigorous definitions that focus on requirements and events, to then specify these for tree-based formalisms in Sec. 6. Furthermore, we find that one of the dependencies — i.e., mutual reinforcement — could be better understood by defining reinforcement first, as it appears not always be a bi-directional interaction. Encouraging more work in translating these definitions for specific categories of formalisms could further narrow this gap. A first practical step might arise from the analysis of real-life case studies, in settings where safety-security interactions are already present.

**Finding #4: No novel metrics were proposed.** No novel metrics were introduced to quantify safety-security interactions. Again, classical metrics were studied, such as the mean time to failure and attacker success probabilities. These metrics remain anchored to typical safety-specific or security-specific formalisms, and are subsequently merged as modelling constructs are. To address this gap, trade offs between safety and security, e.g., through Pareto analysis, could be studied. Trade offs are — in fact — natural to analyze in these contexts and do not necessitate novel analysis methods.

| Formalism | Metric/Analysis | Size |
|---|---|---|
| **FT/AT** | Probability | 8 #Gates |
| **CFTs** | Probability | 8 #Components |
| **AFTs** | Probability, time and cost of attacks | 24, 6, 5 #Gates |
| **SEFTs** | Probability, Mean Time Between Attacks | 6 #Components |
| **FACT Graphs** | Probability | 9 #Gates |
| **BDMPs** | Mean Time to Success, probability of success | 19 #Gates |
| **ATBTs** | Risk level evaluation, trade offs analysis | 12 #Controls |
| **BNs** | Mean Time to Failure, conditional independence | 16 #Nodes |
| **THO Framework** | Likelihood of an attack | / |
| **STAMP** | Potential hazards and undesired behaviours | 6, 10 #Components |
| **SysML** | States that violate properties/requirements | 5 #Blocks |
| **ALLOY** | Prove properties under given assumptions | 18 #Input/Output ports |
| **Event-B** | Prove properties under given assumptions | 6 #Components in safety goal |
| **AADL** | Mean Time Between Failures | 9 #Gates (corresp. FT) |

Table 4: Comparison of case study complexity per analyzed formalism.

**Finding #5: No large case studies were carried out.** In general, papers analyzed in this work present relatively small examples for illustrative purposes (see Table 4). A notable exception is the medium-size, but realistic, pipeline case study in [51] and [54]. Another notable exception is the electrical grid case study in [20]. This is remarkable, because for safety and security separately, such numerous large case studies exist, e.g., [17, 34, 16, 37, 12]. Conducting a thorough case study analysis in a realistic setting could shed light on how safety and security practically interact, on how one could formally define these interactions in a specific context and could help gather more insight on how safety and security can be jointly analyzed.

**Finding #6: Diverse formalisms model different safety-security interactions.** As shown in Table 2 and in Sec. 9, AFTs, BDMPs, BNs, Event-B and ALLOY are the only formalisms that can model all four safety-security dependencies. CFTs and SEFTs can model them provided with extensions/with some limitations, which we suggested in the respective sections, e.g., CFTs need a NOT-gate to express antagonism while SEFTs can only express directional conditional dependency, from security to safety.

## 11. Conclusion and Future Work

In the introduction of this paper, we posed several research questions. In this section, we look back at these questions, offer a summary of answers and some pointers to valuable sections in the paper, while proposing possible lines of research for future work. (**Q1**) How expressive are these formalisms? (**Q2**) Which modeling constructs exist to model the interactions between safety and security? (**Q3**) Which analyses do these formalisms enable? (**Q4**) How do these formalisms compare on industrial case studies and (**Q5**) What are the gaps and the findings? What would be desirable extensions?

**In summary:**

(**A1**) Our paper provides a thorough comparison of the analyzed formalisms. For each formalism, we highlight its expressiveness, i.e., their ability to model safety/security interactions, in the appropriate section and we conduct a thorough comparison between them in Sec. 9. Out of the 14 formalisms analyzed, AFTs, BDMPs, BNs, Event-B and ALLOY are the only 5 formalisms that can model all four safety-security dependencies.

(**A2**) As shown, none of the formalisms present *specific constructs* for safety-security interdependencies: constructs are acquired from safety-only or security-only frameworks and joined afterwards, following different strategies. The same holds for *metrics*: none of those is specific for safety-security interactions. We reflect on reasons behind this phenomenon in Sec. 10.

(**A3**) Studied formalisms enable qualitative and/or quantitative analyses: details are provided and highlighted in each section and summarized in Table 3.

(**A4**) As highlighted in Sec. 10 and to the best of our knowledge, safety-security interactions are studied inside limited scenarios and large industrial case studies are still missing.

(**A5**) We tackle gaps in each section and highlight them while analyzing formalisms. Moreover, we summarize our general findings in Sec. 10. Further extensions could be considered and are suggested in appropriate sections, e.g., for every FT/AT-based formalism, one could employ two roots to account for different safety and security TLEs, similarly to what BNs do.

**Future work.** More rigorous definitions of safety-security dependencies are needed, to account for: 1. *Directionality.* Are safety and security directional or bi-directional and from which direction do they flow? 2. *Intensity.* For a quantifiable co-analysis, intensity of these interaction has to be considered. 3. *Nature of the interaction.* For each of the possible interactions, from reinforcement, to dependency or antagonism, accounting for the positive or negative impact of such an interaction is fundamental. Moreover, conditional dependencies like the one showed in Fig. 4a raise the question on who is responsible for depending actions when safety and security are heavily dependent. Game theoretical frameworks could be deployed to analyze this open issue. The visualization of safety/security interdependencies should also be considered to ease readability of complex models. The most recent update of the Alloy tool has introduced more capabilities than explored in the current literature, exploring how the new temporal modeling capabilities can be exploited for safety-security modeling would be of interest as well.